\documentclass[useAMS,usenatbib,onecolumn]{mnras}
\usepackage{amsmath,bm}
\usepackage{dcolumn}
\usepackage[utf8]{inputenc}
\usepackage{graphics,graphicx}
\usepackage{float}
\usepackage{threeparttable}
\usepackage{url}
\usepackage{epstopdf}

\usepackage{longtable}

\usepackage{color}

\newcommand{\bra}[1]{\langle #1|}
\newcommand{\ket}[1]{|#1\rangle}

\newcommand{\HTO}{H$_3$O$^+$}

\newcommand{\Dh}[1]{${\mathcal D}_{#1{\rm h}}$}

\newcommand{\cm}{cm$^{-1}$}
\newcommand{\um}{$\mu$m}
\newcommand{\ai}{\textit{ab initio}}

\newcommand{\lin}{^{\rm lin}}
\newcommand{\rme}{_{\rm e}}

\graphicspath{{eps/}}

\title[ExoMol line lists -- XL. Hydronium]{ExoMol line lists -- XL. Ro-vibrational molecular line list for the hydronium ion (H$_3$O$^+$)}

\date{\today}

\author[Yurchenko et al.]{S. N. Yurchenko$^1$, Jonathan Tennyson$^1$\thanks{The corresponding author: j.tennyson@ucl.ac.uk}, Steve Miller$^1$, V.V. Melnikov$^2$, \newauthor J. O'Donoghue$^3$, L. Moore$^4$ \vspace*{4mm}\\
$^1$ Department of Physics and Astronomy, University College London, Gower Street, WC1E 6BT London, UK\\
$^2$ Siberian Institute of Physics \& Technology, Tomsk State University, Tomsk, 634050 Russia\\
$^3$ JAXA Institute of Space and Astronautical Science, Japan\\
$^4$ Center for Space Physics, Boston University, Boston, MA, USA}

\begin{document}

\label{firstpage}

\maketitle

\begin{abstract}

A new line list for hydronium (H$_3$$^{16}$O$^+$) is computed. The line list is based on a new  \text{ab initio} dipole moment surface (CCSD(T)/aug-cc-pVQZ) and a new empirical potential energy surface (PES). The empirical PES of H$_3$O$^+$ was obtained by refining an \textit{ab initio} surface  through a global fit to the experimentally determined ro-vibrational energies collected from the literature covering the ground, $\nu_1^{\pm}$, $\nu_2^{\pm}$, $2\nu_2^{\pm}$, $\nu_3^{\pm}$ and $\nu_4^{\pm}$ vibrational states. The line list covers the wavenumber range up  to 10~000~cm$^{-1}$ (wavelengths $>1 \mu$m) and should be complete for temperatures up to $T=1500$~K. This is the first comprehensive line list for H$_3$O$^+$ with extensive wavenumber coverage and accurate transitional probabilities. Prospects of detection of hydronium in spectra of solar system giant planets as well as exoplanets are discussed.  The eXeL line list is publicly available from the ExoMol and CDS databases.


\end{abstract}

\begin{keywords}
molecular data;  astronomical data bases: miscellaneous; planets and satellites: atmospheres;  planets and satellites: gaseous planets; Physical data and processes.
\end{keywords}

\section{Introduction}

Hydronium and its isotopologues play an important role in planetary and (inter)stellar chemistry \citep{76DaBlxx,00JeBiSa.H3O+,01GoCexx.H3O+,10GeLuBl.H3O+,
12HoKaNe.H3O+,13GoFiBr.H3O+,15SaVeAg.H3O+,15InNeGe.H3O+,18TrReKo.H3O+,19MaAgBo.H3O+}.
These ions are found to exist abundantly in both diffuse and dense molecular clouds \citep{86HoChHe.H3O+,86WoBoBo.H3O+,91WoMaTu.H3O+,92PhVaKe.H3O+,96TiNiPo,01GoCexx.H3O+,10GeLuBl.H3O+,13GoFiBr.H3O+,12HoKaNe.H3O+,15InNeGe.H3O+}  as well as in comae  \citep{09RuHaGo.comets}. \HTO\ is an indicator of of the presence of water and can be used to estimate H$_2$O abundances when the direct detection is unfeasible \citep{92PhVaKe.H3O+,15RoDaxx.H3O+}.
\HTO\ was detected in comets Hale-Bopp and Halley \citep{97Rauer,97MeGaBe,97LiMeBe,86BaAlBu}. Observations of \HTO\ are one of the approaches to establish interstellar concentrations of H$_2$O.
Dissociative recombination of \HTO\ with electrons  is thought to be the main source for the synthesis of water dense interstellar clouds  \citep{88MiDeMc.H3O+,91WoMaTu.H3O+,96AnHeKe.H3O+}
and  may lead to formation of a population of vibrationally hot  water in comets \citep{jt402}.

\HTO\ is expected to exist in a wide variety of environments, such as diffuse interstellar clouds,  at very low temperatures or, for example, the atmospheres of giant planets \citep{18MoCrMu}, brown dwarfs and cool stars which are significantly hotter.
Recent laboratory experiments by  \citet{jt793} suggest the \HTO\ is likely to be both the dominant and the most easily observed molecular ions in sub Neptune exoplanets; \citet{jt793} also suggest that \HTO\ should be observable by forthcoming exoplanet characterisation space missions, such a detection would require a reliable  line list for hot \HTO. \citet{19HeRixx} suggest that  H$_3$O$^+$ should be detectable  in free-floating brown dwarfs and super-hot giants.

Dissociative recombination of hydronium H$_3$O$^{+}$ has been extensively studied in ion storage rings by \cite{96AnHeKe.H3O+, 00NeKhRo.H3O+, 00JeBiSa.H3O+, 10BuStMe.H3O+,10NoBuSt.H3O+}. The sensitivity of the Hydronuim spectrum to variations of the electron-to-proton mass ratio was studied by \citet{10KoLexx.H3O+,15OwYuPo.H3O+}.



The \HTO\ ion is destroyed primary by electrons and ammonia \citep{19HeRixx}.
\HTO\ is one of the species used in the spectroscopic breath analysis \citep{19SpSpSm.H3O+}.

Hydronium (\HTO) is a pyramidal molecule characterized by an umbrella motion with a very low barrier to the planarity of  around 650.9~\cm \citep{04RaNoVa.H3O+}. As the result vibrational ground state is split due to tunnelling by  55.35~\cm\ \citep{99TaOkxx.H3O+}, significantly more than in the isoelectronic ammonia molecule.

Theoretical studies of structure, inversion barrier and ro-vibrational energy levels of \HTO\ were carried out by
\citet{73LiDyxx.H3O+, 80FeHaxx.H3O+, 82SpBuxx.H3O+, 83BoRoRe.H3O+, 86LiOkSe.H3O+, 05YuBuJe.H3O+}.
The electronic structure of hydronium and hydronium-water clusters were studied by \citet{02ErSoDo.H3O+}.

Accurate \textit{ab initio} studies include works by \citet{00ChJuGe.H3O+,03RaMiHaa.H3O+,03HuCaBo.H3O+,04RaNoVa.H3O+,13MaXiSa.H3O+,15OwYuPo.H3O+,16YuBoxx.H3O+}, where potential energy and dipole moment surfaces of \HTO\ were computed using high levels of theory.

Chemistry of \HTO\ was also a subject of numerous studies
\citep{12HoKaNe.H3O+, 15RoDaxx.H3O+, 16CrBeHo.H3O+}.
The influence of the liquid environment on the spectroscopic properties of \HTO\ was studied by \citet{16TaLiCh.H3O+}.

Experimental  data on the high-resolution line positions of \HTO\ were collected by \citep{09YuDrPe.H3O+} from a large set of  high-resolution spectroscopy studies 
\citep{83BeGuPf.H3O+, 85BoDeDe.H3O+, 84DaHaJo.H3O+, 84BuAmSp.H3O+,
85BeSaxx.H3O+, 85LiOkxx.H3O+, 85PlHeEr.H3Op, 86DaJoHa.H3O+,
87GrPoSa.H3O+, 88HaLiOk.H3O+,
88VeTeMe.H3O+, 90OkYeMy.H3O+, 90PeNeOw.H3O+,
91HoPuOk.H3O+, 97UyWhOk.H3O+, 99ArOzSa.H3O+,
06DoNexx.H3O+, 08FuFaxx.H3O+, 09YuDrPe.H3O+, 10MuDoNe.H3O+,
12PeWeBe.H3O+, 84LeDexx.H3O+, 85SeBuDa.H3O+,
84HaOkxx.H3O+, 87StAdUr.H3O+, 89VeVeTe.H3O+, 99TaOkxx.H3O+,
05StSaxx.H3O+, 07ZhWaLi.H3O+}.
\cite{09YuDrPe.H3O+} used these data in a global spectroscopic analysis with the \textsc{SPFIT}/\textsc{SPCAT} effective Hamiltonian approach \citep{SPFIT}, together the \HTO\ line positions from NASA JPL \citep{JPL}. They have computed a set of empirical energies of \HTO\ for $J=0,\ldots, 20$  for the ground as well as the $\nu_1^{\pm}$, $\nu_2^{\pm}$, $2\nu_2^{\pm}$, $\nu_3^{\pm}$ and $\nu_4^{\pm}$ vibrational states. We use these energies to refine our spectroscopic data.

In our recent work on \HTO\ \citep{jt660} we computed a low temperature line list for all main isotopologues of  \HTO\ using the \ai\ potential energy surface (PES) and dipole moment surface (DMS) of \citet{15OwYuPo.H3O+}  combined with accurate variational nuclear motion calculations using  TROVE \citep{TROVE}, where accurate lifetimes of these ions were reported.
Here we present a new hot line list for the main isotopologue of \HTO\ generated using a new \ai\ DMS (CCSD(T)/aug-cc-pVQZ), a new spectroscopic PES, and the program TROVE. This work is performed as part of the ExoMol project which provides
 molecular line lists for exoplanet and other atmospheres \citep{jt528}.

\section{Potential energy surface}

The PES was represented by a Taylor-type expansion of 8$^{\rm th}$ order
\begin{equation}
\label{e:PEF}
V(r_1,r_2,r_3,\alpha_{12},\alpha_{13},\alpha_{23}) = \sum_{i_1,i_2,i_3,i_2,i_3,i_6} f_{i_1,i_2,i_3,i_2,i_3,i_6}
\chi_1^{i_1} \chi_2^{i_2} \chi_3^{i_3} \chi_4^{i_4} \chi_5^{i_5} \chi_6^{i_6}
\end{equation}
using the following definition of the expansion coordinates:
\begin{eqnarray}
\label{eq:icoord1}
\chi_k&=&1-\exp\left[-a(r_k-r\rme)\right], \quad (k=1,2,3), \\
\label{eq:icoord2}
\chi_{4}&=&\left(2\alpha_{23}-\alpha_{13}-\alpha_{12}\right)/\sqrt{6}, \\
\chi_{5}&=&\left(\alpha_{13}-\alpha_{12}\right)/\sqrt{2}, \\
\label{eq:sin:brho}
  \chi_6 &= & \sin {\bar \rho} \, =\, \frac{2}{\sqrt{3}} \,\sin \left[
     (\alpha_{23} + \alpha_{13} + \alpha_{12}) /6\right] \hbox{,}
\end{eqnarray}
where  $r_1$, $r_2$, $r_3$ are the three bond lengths,  $\alpha_{12}$, $\alpha_{13}$ and $\alpha_{23}$ are the three inter-bond angles and umbrella mode $\bar{\rho}$ is an angle between a symmetric $C_3$ axis and molecular bonds $r_i$ for a reference structure defined by the mean angle $\bar\alpha$:
$$
\bar\alpha = \frac{1}{3} (\alpha_{23}+ \alpha_{13}+ \alpha_{12}).
$$
The potential energy function $V(r_1,r_2,r_3,\alpha_{12},\alpha_{13},\alpha_{23})$ of \HTO\  must be fully symmetric and transform as  $A_1'$ of \Dh{3}(M). Therefore the potential parameters $f_{i_1,i_2,i_3,i_4,i_5,i_6}$ are related through the corresponding symmetry properties of $A_1'$. Here we use the same symmetry-adapted expansion of the potential energy function as developed for ammonia in \citet{05YuZhLi.NH3}.

\section{Dipole moment surface} \label{DMS}

A new  high-level \ai\ DMS OF \HTO\ was computed with MOLPRO \citep{MOLPRO} using the CCSD(T)/aug-cc-pVQZ level of theory (coupled cluster with all single and double excitations and a perturbational estimate of connected triple excitations) and the augmented correlation consistent quadruple zeta basis set~\citep{89Dunning.ai,92KeDuHa.ai,93WoDuxx.ai,01DuPeWi.ai}, in the frozen core approximation. The three components of electronically averaged dipole moment $\bar{\bf{\mu}}$ of \HTO\ were obtained as finite difference derivatives with respect to a weak (0.002~atomic units) external field on a grid of 26~271 geometries covering the energy range up to $hc\cdot$23~000 \cm\ with the bond lengths and bond angles varying as 0.96--1.3~\AA\ and 70--125$^{\circ}$, respectively. These three DMSs were then represented analytically using the symmetrized
molecular bond (SMB) representation of \citet{jt466}. In this representation the dipole moment vector $\bar{\bf{\mu}}$ is given by
symmetrized projections onto the molecular bonds with the dipole moment components ($\overline{\mu}_{A_2''}$,$\overline{\mu}_{E_a'}$,$\overline{\mu}_{E_b'}$) in
the molecule fixed axis system given by  6$^{\rm th}$ order polynomial expansions
\begin{equation}
\bar{\mu}_\Gamma(r_1,r_2,r_3,\alpha_{12},\alpha_{13},\alpha_{23})=  \sum_{i_1,i_2,i_3,i_2,i_3,i_6} \mu^{\Gamma}_{i_1,i_2,i_3,i_2,i_3,i_6}
\zeta_1^{i_1} \zeta_2^{i_2} \zeta_3^{i_3} \zeta_4^{i_4} \zeta_5^{i_5} \zeta_6^{i_6},
\end{equation}
where $\Gamma=A''_2, E'_a$ and $E'_b$ are the irreducible components of \Dh{3}(M) \citep{98BuJe.method},
\begin{eqnarray}
\label{eq:mu:icoord1}
\zeta_k &=& (r_k - r_{\rm ref}) \exp\left[-(r_k - r_{\rm ref})\right], \quad (k=1,2,3) \\
\label{eq:mu:icoord2}
\zeta_{4}&=&\left(2\alpha_{23}-\alpha_{13}-\alpha_{12}\right)\sqrt{6}, \\
\zeta_{5}&=&\left(\alpha_{13}-\alpha_{12}\right)/\sqrt{2}, \\
\zeta_6 &=& \sin {\bar \rho},
\end{eqnarray}
$\mu^{\Gamma(s)}_{ij...}$ are the expansion parameters, $r_{\rm ref}$ is a reference bond length used as an expansion center
and $\sin\bar\rho$ is the same as  in Eq.~(\ref{eq:sin:brho}). The dipole moment expansion parameters are obtained in a least-squares fit to the \ai\ values.
The dipole moment components ($\overline{\mu}_{E'_a},\overline{\mu}_{E'_b}$) are transformed as linear combinations of each other according with the  irreducible representation $E'$ of \Dh{3}(M). In the symmetry-adapted form of \citet{jt466} used here, the  parameters $(\mu^{E_A(s)}_{ij...},\mu^{E_b(s)}_{ij...})$ are shared between these two components and  thus must be fit together, while  $\mu^{A_1(s)}_{ij...}$ are obtained separately (see \citet{jt466}).

Since the dipole moments of ions depend on the origin of the coordinate, for variational calculations the \ai\ DMS of \HTO\ had to be transformed from the coordinate system centered on the oxygen atom used in the MOLPRO calculations to the center of mass (CM) $\bar\mu_{\alpha}^{\rm (CM)}$ as follows
$$
\bar\mu_{\alpha}^{\rm (CM)} = \bar\mu_{\alpha}^{\rm (O)} - R_{\alpha}^{\rm (CM)} Q.
$$
where $R_{\alpha}^{\rm (CM)}$ are the Cartesian coordinates of the center of mass of \HTO\ in the old coordinate system and $Q = 4.803206798$~Debye$/$\AA\ is the charge of the ion in Debye$/$\AA.

The final \ai\ dipole moment functions (DMF) required {221} parameters and reproduced the \ai\ data with an root-mean-squares (rms) error  of 0.014~D for geometries with energies up to $hc\cdot$24~000 \cm\ and 0.00036~D for geometries with energies up to $hc\cdot$12 000~\cm. The \ai\ DMF of \HTO\ is included in the supplementary material as a Fortran 90 routine.

Using our \ai\ (CM) dipole moment and the TROVE vibrational eigenfunctions (see below) we obtained a transition dipole moment for the  inversion $0^- \leftrightarrow 0^+$ band of 1.438~D, which coincides with the \ai\ value of \citet{83BoRoRe.H3O+}  adopted  by the CDMS database \citep{CDMS} as the ground state permanent  dipole moment.


\section{TROVE specifications}

 Owing to the low barrier, the ro-vibrational motion of \HTO\ is described by the \Dh{3}(M) molecular symmetry group \citep{98BuJe.method}.
For this work we use a  similar setup to that adopted by  \citet{15OwYuPo.H3O+} and \citet{jt660}. The kinetic energy operator was constructed as a 6th order expansion in terms five rectilinear linearized coordinates $\xi_{i}\lin$ =  $\{\Delta r_1\lin, \Delta r_2\lin,\Delta r_3\lin, S_A\lin,S_B\lin\}$  around a non-rigid configuration following the Hougen-Bunker-Johns approach \citep{70HoBuJo} as implemented in TROVE. These coordinates are linearized versions of the geometrically defined coordinates $\xi_{i}$ ($i=1,\ldots,5$), constructed from the valence coordinates $r_1$, $r_2$, $r_3$, $\alpha_{12}$, $\alpha_{13}$ and $\alpha_{23}$ as follows:
\begin{align}\label{e:r1}
  \xi_1 &= r_{1}-r\rme, \\
  \xi_2 &= r_{2}-r\rme, \\
  \xi_3 &= r_{3}-r\rme, \\
  \xi_4 &= \frac{1}{\sqrt{6}} \left( 2\alpha_{23} - \alpha_{13} - \alpha_{12}   \right), \\
  \xi_5 &= \frac{1}{\sqrt{2}} \left( \alpha_{13} - \alpha_{12}   \right),
\end{align}
where $r\rme$ is the  bond length of \HTO.
The 6th, inversion coordinate $\tau$ is defined as an angle between any of the bonds and their trisector ${\bf n}$.

The 1D primitive vibrational basis functions $\phi_{v_i}(\xi\lin_i)$ ($i=1\ldots 5$) and $\phi_6(\tau)$ were defined as follows. We used a numerically generated based set for the stretching modes using the Numerov-Cooley approach \citep{24Numerov.method,61Cooley.method}, where 1D stretching Schr\"{o}dinger equations were solved on a grid of 2000 points $r_i\lin$ ranging from  0.4 to 2.0 \AA.  1D Harmonic oscillator functions were used to form the bending basis sets for $\xi_4$ and $\xi_5$. The inversion basis set was also constructed using the Numerov-Cooley approach on a grid of 8000 $\tau$ points ranging from $-55^{\circ}$ to $55^{\circ}$. The stretching primitive basis functions $\phi_{v_i}(\xi\lin_i)$ ($i=1,2,3$) were selected to cover $v_i=0\ldots 9$, while the excitations of the bending and inversion basis functions extended to $v_i = 36$ ($i=4,5,6$). 1D Hamiltonians for each mode used for the stretching and inversion 1D problems were constructed from the 6D Hamiltonian by setting all other coordinates to their equilibrium values.

TROVE uses a two-step basis set optimization scheme designed to improve the sum-of-product form of the vibrational basis set, see \citet{17YuYaOv.methods}. At step 1, three sets of reduced Hamiltonian problems are solved for (i) the sub-group of 3 stretching modes, (ii) the sub-group of two bending modes and (iii) the inversion mode. These sub-groups are organized to form symmetry independent modes so that the eigenfunctions of the corresponding reduced Hamiltonians can be symmetrized and classified according with \Dh{3}(M) using the TROVE automatic symmetrization procedure \citep{17YuYaOv.methods}. The three reduced Hamiltonians are given by
\begin{equation}
\begin{split}
  \hat{H}_{\rm str}^{(1)}(\xi_1\lin,\xi_2\lin,\xi_3\lin) &= \bra{0_4} \bra{0_5} \bra{0_6} \hat{H}^{\rm 6D} \ket{0_6} \ket{0_5} \ket{0_4},  \\
\hat{H}_{\rm bnd}^{(2)}(\xi_4\lin,\xi_5\lin) &= \bra{0_1} \bra{0_2} \bra{0_3} \bra{0_6} \hat{H}^{\rm 6D} \ket{0_6} \ket{0_3} \ket{0_2} \ket{0_1},\\
\hat{H}_{\rm inv}^{(3)}(\rho) &= \bra{0_1} \bra{0_2} \bra{0_3}\bra{0_4} \bra{0_5} \hat{H}^{\rm 6D} \ket{0_5} \ket{0_4} \ket{0_3} \ket{0_2} \ket{0_1}.
\end{split}
\end{equation}

The final vibrational basis set is a direct product of the corresponding three  basis sets. This product is contracted using the following polyad-number condition:
$$
P  = 4 (v_1+v_2+v_3) + 2(v_4 + v_5) + v_6 \le  P_{\rm max} = 36.
$$
Once the vibrational part is solved, the eigenfunctions of the $J=0$ Hamiltonian are contracted again using the condition $\tilde{E}_i \le 20000$~\cm\  and used to form our final ro-vibrational basis as a product with the rotational basis functions. The latter are chosen as the symmetrized rigid rotors wavefunctions (see \citet{TROVE}). The vibrational ($J=0$) contracted basis set comprised 9134 functions. Using this basis set, for each value of $J$ from 0 to 40, four symmetry-adapted Hamiltonian matrices ($A'_2$, $E'$, $A''_2$ and $E''$) were constructed and diagonalized to obtained energies and eigenvectors up to $\tilde{E} = 18\,000$~\cm. The $A'_1$ and $A''_1$ eigenfunctions correspond to nonphysical representations with the nuclear statistical weights $g_{\rm ns}$ equal zero. The $A_2$ and  $E$-type symmetries have nuclear statistical weights of 4 and 2, respectively.

The ro-vibrational states of \HTO\ were assigned the TROVE  quantum numbers (QN) $J$, $K$,  $v_1$, $v_2$, \ldots $v_6$ as well the symmetry labels $\Gamma_{\rm vib}$ and $\Gamma_{\rm rot}$ using the largest eigen-contribution from   from the primitive or contracted basis functions. For spectroscopic applications, we also
provide the standard  normal mode quantum numbers  $n_1$, $n_2$, $n_3$, $l_3$, $n_4$, $l_4$  reconstructed by correlating them  to $v_1$, $v_2$, \ldots $v_6$. Here $n_1$ is the symmetric ($A'_1$) stretching QN, respectively; $n_2$ is the inversion QN  ($A''_2$); $n_3$ and $n_4$ are asymmetric ($E'$) stretching and bending QNs, respectively and $l_3$ and $l_4$ are the corresponding vibrational angular momentum QNs. The latter satisfy the standard 2D isotropic oscillator conditions \citep{98BuJexx}
$$
l_i = v_i, v_i-2, \ldots, 0 (1).
$$
These QNs were estimated as eigenvalues of the vibrational angular momentum operator squared $L_z^2$ on the primitive bending basis functions $\phi_{v_3}(\xi\lin_3)$  and $\phi_{v_4}(\xi\lin_4)$ following the methodology described in \citet{jt771}. 

In order to improve the quality of the calculated energies and the line list positions of \HTO, an empirical PES of \HTO\ has been constructed by refining \ai\ PES of \HTO\ of \citet{15OwYuPo.H3O+} to the available
laboratory spectroscopic data. In this fits, we used the empirical \HTO\ energies  collected by  \citet{09YuDrPe.H3O+}, which were constructed as a global fit to the experimental line positions from the literature (see Introduction for the detailed references) using \textsc{SPFIT}/\textsc{SPCAT} \citep{SPFIT}. Their analysis covered the pure rotational as well as the $\nu_1$, $\nu_2$, $2\nu_2$, $\nu_3$ and $\nu_4$ vibrational states.  Our fitting set comprised of energies for $J=0,1,2,3,4,6,8,10,16$ and is illustrated in Table~\ref{t:obs-cacl}, which also shows the quality of the energies obtained with the refined PES. A few states are found with  large or very large residuals ($>20$ \cm), which we believe are outliers of the \textsc{SPFIT}/\textsc{SPCAT} analysis.

Due the limited coverage of the experimental information, the refined PES is still largely based on the initial \ai\ surface thus affecting the accuracy of the fit and as well as the quality of the energies and line positions extrapolated outside the experimental set, especially for higher excitations corresponding to large distortions of PES.  Another source of the inaccuracy is from the non-exact kinetic energy operator (KEO) formalism used in the variational calculations (see \citet{TROVE}) mostly  affecting energies at high $J$s. The KEO errors are usually much  smaller (10--100 times) than the errors associated with the \ai\ character of PES.   Even with these caveats we  believe that our results represent a significant improvement to the existing knowledge of the hydronium spectroscopy especially at higher vibrational or rotational excitations. 

The  potential parameters $f_{i_1,i_2,i_3,i_2,i_3,i_6}$ from Eq.~\eqref{e:PEF} representing the  refined potential energy function of \HTO\   $V(r_1,r_2,r_3,\alpha_{12},\alpha_{13},\alpha_{23})$ are given in the  supplementary material together with a Fortran program. 
It is expressed in terms  of the valence coordinates $r_i$ and $\alpha_{jk}$ independent from the special coordinate choice used in TROVE and thus can be used with any other programs. It should be noted however that because of the approximations used in TROVE  (non-exact KEO, incomplete basis set etc, linearization of the valence coordinates in the the representation of PES), the ro-vibrational energies obtained using our refined PES are expected to be somewhat different from ours.

The ro-vibrational energies and wavefunctions were computed variationally using the refined PES for $J=0\ldots 40$. The transitional intensities (Einstein~A coefficients) were generated with our GPU code GAIN-MPI \citep{jt653} in conjunction with the \ai\ DMS described above.


\begin{table*}
\caption{\label{t:obs-cacl} A comparison of the calculated energy (Calc.) term values (\cm) of \HTO\ with the experimental or empirically derived (Obs.) term values and band centers (\cm) for $J=0,1,2$.
The complete table of the fitting set ($J=0,1,2,3,4,6,8,10,16$) in given in supplementary material.
}
\centering
\begin{tabular}{rllrrrcrllrrr}
\hline\hline
$J$         &  $\Gamma$      &  State      &  Obs.          &   Calc.     &  Obs.-Calc. && $J$         &  $\Gamma$      &  State      &  Obs.          &   Calc.     &  Obs.-Calc.    \\
\hline
       0  &  $A_1'$   &$  g.s.       $&       0.0000  &         0.0000 &$    0.0000  $&&         2  &  $A_2'$   &$  \nu_4^+    $&    1669.0705  &      1669.0055 &$    0.0649  $\\
       0  &  $A_1'$   &$  \nu_2^+    $&     581.1768  &       581.1194 &$    0.0574  $&&         2  &  $A_2'$   &$  \nu_1^+    $&    3584.6243  &      3584.5699 &$    0.0544  $\\
       0  &  $A_1'$   &$  2\nu_2^+   $&    1475.8400  &      1476.6341 &$   -0.7941  $&&         2  &  $A_2'$   &$  \nu_1^-    $&    3634.4900  &      3634.5577 &$   -0.0677  $\\
       0  &  $A_1'$   &$  \nu_1^+    $&    3445.0024  &      3445.1247 &$   -0.1223  $&&         2  &  $E'$     &$  g.s.       $&      47.0775  &        47.0957 &$   -0.0182  $\\
       0  &  $E'$     &$  \nu_4^+    $&    1626.0202  &      1625.9707 &$    0.0494  $&&         2  &  $E'$     &$  \nu_2^-    $&     116.8069  &       116.8767 &$   -0.0698  $\\
       0  &  $E'$     &$  \nu_3^+    $&    3536.0364  &      3536.0017 &$    0.0347  $&&         2  &  $E'$     &$  \nu_2^+    $&     627.9416  &       627.9046 &$    0.0370  $\\
       0  &  $A_2''$  &$  \nu_2^-    $&      55.3275  &        55.4027 &$   -0.0752  $&&         2  &  $E'$     &$  2\nu_2^-   $&    1014.1295  &      1014.1451 &$   -0.0156  $\\
       0  &  $A_2''$  &$  2\nu_2^-   $&     954.3777  &       954.3953 &$   -0.0175  $&&         2  &  $E'$     &$  \nu_4^+    $&    1695.4812  &      1695.4250 &$    0.0561  $\\
       0  &  $A_2''$  &$  \nu_1^-    $&    3491.1533  &      3491.3405 &$   -0.1871  $&&         2  &  $E'$     &$  \nu_4^-    $&    1754.2401  &      1754.3021 &$   -0.0619  $\\
       0  &  $E''$    &$  \nu_4^-    $&    1693.9311  &      1694.0408 &$   -0.1096  $&&         2  &  $E'$     &$  \nu_1^+    $&    3491.5374  &      3491.6390 &$   -0.1017  $\\
       0  &  $E''$    &$  \nu_3^-    $&    3574.7899  &      3574.7419 &$    0.0481  $&&         2  &  $E'$     &$  \nu_1^-    $&    3551.8320  &      3551.9705 &$   -0.1386  $\\
       1  &  $A_2'$   &$  g.s.       $&      22.4811  &        22.5019 &$   -0.0208  $&&         2  &  $E'$     &$  \nu_3^+    $&    3580.6437  &      3580.6524 &$   -0.0088  $\\
       1  &  $A_2'$   &$  \nu_2^+    $&     603.5178  &       603.4797 &$    0.0381  $&&         2  &  $E'$     &$  \nu_1^+    $&    3602.1477  &      3602.1407 &$    0.0070  $\\
       1  &  $A_2'$   &$  \nu_4^-    $&    1714.4010  &      1714.5415 &$   -0.1405  $&&         2  &  $E'$     &$  \nu_3^-    $&    3636.3740  &      3636.3380 &$    0.0361  $\\
       1  &  $A_2'$   &$  \nu_1^+    $&    3467.1385  &      3467.2640 &$   -0.1255  $&&         2  &  $A_2''$  &$  \nu_2^-    $&     121.6192  &       121.6888 &$   -0.0696  $\\
       1  &  $A_2'$   &$  \nu_3^-    $&    3590.7923  &      3590.7246 &$    0.0677  $&&         2  &  $A_2''$  &$  2\nu_2^-   $&    1018.5442  &      1018.5605 &$   -0.0164  $\\
       1  &  $E'$     &$  \nu_2^-    $&      72.6131  &        72.6866 &$   -0.0735  $&&         2  &  $A_2''$  &$  \nu_4^+    $&    1693.3494  &      1693.2513 &$    0.0982  $\\
       1  &  $E'$     &$  2\nu_2^-   $&     971.3537  &       971.3701 &$   -0.0164  $&&         2  &  $A_2''$  &$  \nu_4^-    $&    1736.1900  &      1736.2166 &$   -0.0266  $\\
       1  &  $E'$     &$  \nu_4^+    $&    1649.1761  &      1649.1168 &$    0.0593  $&&         2  &  $A_2''$  &$  \nu_1^-    $&    3556.5593  &      3556.6933 &$   -0.1340  $\\
       1  &  $E'$     &$  \nu_4^-    $&    1708.8804  &      1708.9578 &$   -0.0774  $&&         2  &  $A_2''$  &$  \nu_1^+    $&    3596.1048  &      3596.0093 &$    0.0955  $\\
       1  &  $E'$     &$  \nu_1^-    $&    3508.2267  &      3508.4025 &$   -0.1758  $&&         2  &  $A_2''$  &$  \nu_1^-    $&    3623.3991  &      3623.3456 &$    0.0535  $\\
       1  &  $E'$     &$  \nu_1^+    $&    3558.0829  &      3558.0571 &$    0.0258  $&&         2  &  $E''$    &$  g.s.       $&      62.3662  &        62.3838 &$   -0.0176  $\\
       1  &  $E'$     &$  \nu_3^-    $&    3592.8914  &      3592.8400 &$    0.0513  $&&         2  &  $E''$    &$  \nu_2^-    $&     102.3574  &       102.4282 &$   -0.0709  $\\
       1  &  $A_2''$  &$  \nu_4^+    $&    1645.4437  &      1645.4003 &$    0.0434  $&&         2  &  $E''$    &$  \nu_2^+    $&     643.1580  &       643.1157 &$    0.0423  $\\
       1  &  $A_2''$  &$  \nu_1^+    $&    3552.2612  &      3552.2819 &$   -0.0208  $&&         2  &  $E''$    &$  2\nu_2^-   $&    1000.8832  &      1000.8971 &$   -0.0138  $\\
       1  &  $E''$    &$  g.s.       $&      17.3803  &        17.4012 &$   -0.0209  $&&         2  &  $E''$    &$  \nu_4^+    $&    1687.7512  &      1687.6962 &$    0.0550  $\\
       1  &  $E''$    &$  \nu_2^+    $&     598.4434  &       598.4071 &$    0.0363  $&&         2  &  $E''$    &$  \nu_4^-    $&    1746.3154  &      1746.4770 &$   -0.1616  $\\
       1  &  $E''$    &$  \nu_4^+    $&    1641.4819  &      1641.4223 &$    0.0595  $&&         2  &  $E''$    &$  \nu_4^-    $&    1762.0066  &      1762.1048 &$   -0.0982  $\\
       1  &  $E''$    &$  \nu_4^-    $&    1716.6186  &      1716.7175 &$   -0.0990  $&&         2  &  $E''$    &$  \nu_1^+    $&    3506.4561  &      3506.5479 &$   -0.0918  $\\
       1  &  $E''$    &$  \nu_1^+    $&    3462.1613  &      3462.2898 &$   -0.1285  $&&         2  &  $E''$    &$  \nu_1^-    $&    3537.6298  &      3537.7900 &$   -0.1603  $\\
       1  &  $E''$    &$  \nu_3^+    $&    3554.1949  &      3554.1491 &$    0.0458  $&&         2  &  $E''$    &$  \nu_3^+    $&    3598.2704  &      3598.2424 &$    0.0280  $\\
       1  &  $E''$    &$  \nu_1^-    $&    3596.5379  &      3596.4975 &$    0.0404  $&&         2  &  $E''$    &$  \nu_1^-    $&    3640.0122  &      3639.9883 &$    0.0239  $\\
\hline
\hline
\end{tabular}
\end{table*}

\section{Line list}


The ro-vibrational energies and Einstein~A coefficients were then compiled into a line list eXeL utilizing the two-parts ExoMol format \citep{jt631}, consisting of States and Transitions files.  The line list consists of 1~173~114 states and 2~089~331~073 transitions covering the energy range up to $hc\cdot$18~000~\cm\ and the wavenumber range up to 10~000~\cm\ with the lower energy value limited by $hc\cdot$ 10~000~\cm. The transitions are split into 100 Transition files of 100~\cm\ each. Extracts from the States and Transition files are shown in Tables~\ref{t:states} and ~\ref{t:trans}, illustrating their  structure and quantum numbers. The ro-vibrational states of \HTO\ are assigned with the following quantum numbers: the total angular momentum $J$; the projection of $J$ on the molecular axis $k$; the total, vibrational and rotational symmetries $\Gamma_{\rm tot}$, $G_{\rm vib}$ and $\Gamma_{\rm rot}$ in \Dh{3}(M), respectively; the local mode quantum numbers $v_1$, $v_2$, $v_3$ (stretches), $v_4$, $v_5$ (bends) and $v_6$ (inversion) in accordance with the corresponding vibrational primitive basis functions as described above and the normal mode quantum numbers $n_1$, $n_2$,  $n_3$, $l_3$, $n_4$, $l_4$.

In order to improve the quality of the line list further, the TROVE theoretical energies were replaced by the empirical values from \cite{09YuDrPe.H3O+} where possible.

\begin{table}
{\tt
\footnotesize
\caption{\label{t:states} Extracts from the final states file for the eXeL line.}
\tabcolsep=5pt
\begin{tabular}{rrrrrcrrrrrrcrcrrrrrrrr}
\hline\hline
 & & & & & &  \multicolumn{6}{c}{\rm Normal mode QN} & & \multicolumn{2}{c}{\rm Rot. QN} & &\multicolumn{6}{c}{\rm TROVE QN}\\
\cline{7-12}
\cline{14-15}
\cline{17-22}
$i$ &  \multicolumn{1}{c}{$\tilde{E}$}  & $g_{\rm tot}$  & $J$ & {\rm unc.} &\multicolumn{1}{c}{$\Gamma$}&  $n_1$ &$n_2$ &$n_3$& $l_3$& $n_4$ &$l_4$& \multicolumn{1}{c}{$\Gamma_{\rm vib}$} & $K$ & \multicolumn{1}{c}{$\Gamma_{\rm rot}$} & \multicolumn{1}{c}{$C_i$} & $v_1$ &$v_2$ &$v_3$& $v_4$& $v_5$ &$v_6$ \\
\hline
  3476  &   22.481050  &    12   & 1  &   0.0000  & A2' &    0  &  0  &  0 &   0  &  0  &  0 & A1'  &   0 & A2' &   1.00  &     0  &  0  &  0  &  0  &  0  &  0 \\
  3477  &  603.517820  &    12   & 1  &   0.0014  & A2' &    0  &  1  &  0 &   0  &  0  &  0 & A1'  &   0 & A2' &  -1.00  &     0  &  0  &  0  &  0  &  0  &  2 \\
  3478  & 1497.455621  &    12   & 1  &     0.41  & A2' &    0  &  2  &  0 &   0  &  0  &  0 & A1'  &   0 & A2' &   1.00  &     0  &  0  &  0  &  0  &  0  &  4 \\
  3479  & 1714.401040  &    12   & 1  &   0.0025  & A2' &    0  &  0  &  0 &   0  &  1  &  1 & E"   &   1 & E"  &   1.00  &     0  &  0  &  0  &  0  &  1  &  1 \\
  3480  & 2621.469120  &    12   & 1  &     0.41  & A2' &    0  &  2  &  0 &   0  &  0  &  0 & E"   &   1 & E"  &   1.00  &     0  &  0  &  0  &  0  &  1  &  3 \\
  3481  & 2692.172161  &    12   & 1  &     0.41  & A2' &    0  &  1  &  0 &   0  &  1  &  1 & A1'  &   0 & A2' &  -1.00  &     0  &  0  &  0  &  0  &  0  &  6 \\
  3482  & 3237.598553  &    12   & 1  &     0.41  & A2' &    0  &  1  &  0 &   0  &  1  &  1 & A1'  &   0 & A2' &  -1.00  &     0  &  0  &  0  &  0  &  2  &  0 \\
  3483  & 3336.105179  &    12   & 1  &     0.61  & A2' &    0  &  2  &  0 &   0  &  1  &  1 & E"   &   1 & E"  &  -1.00  &     0  &  0  &  0  &  0  &  2  &  1 \\
  3484  & 3467.138460  &    12   & 1  &   0.0036  & A2' &    0  &  0  &  0 &   0  &  2  &  0 & A1'  &   0 & A2' &   1.00  &     0  &  0  &  1  &  0  &  0  &  0 \\
  3485  & 3590.792340  &    12   & 1  &   0.0012  & A2' &    0  &  0  &  0 &   0  &  2  &  0 & E"   &   1 & E"  &   1.00  &     0  &  1  &  0  &  0  &  0  &  1 \\
  3486  & 3735.190221  &    12   & 1  &     0.41  & A2' &    0  &  0  &  0 &   0  &  2  &  2 & E"   &   1 & E"  &  -1.00  &     0  &  0  &  0  &  0  &  1  &  5 \\
  3487  & 3820.010491  &    12   & 1  &     0.41  & A2' &    0  &  0  &  0 &   0  &  2  &  2 & A1'  &   0 & A2' &   1.00  &     0  &  0  &  0  &  0  &  2  &  2 \\
  3488  & 4049.835413  &    12   & 1  &     0.21  & A2' &    1  &  0  &  0 &   0  &  0  &  0 & A1'  &   0 & A2' &  -0.99  &     0  &  0  &  1  &  0  &  0  &  2 \\
\hline\hline
\end{tabular}
}
\mbox{}\\

{\flushleft
$i$:   State counting number.     \\
$\tilde{E}$: State energy term value in \cm. \\
$g_{\rm tot}$: Total state degeneracy.\\
$J$: Total angular momentum.            \\
unc.: Uncertainty \cm.     \\
$\Gamma$:   Total symmetry index in \Dh{3}(M)\\
$n_1$: Normal mode stretching symmetry ($A'_1$) quantum number. \\
$n_2$: Normal mode inversion ($A''_2$) quantum number. \\
$n_3$: Normal mode stretching asymmetric ($E'$) quantum number. \\
$l_3$: Normal mode stretching angular momentum  quantum number.\\
$n_4$: Normal mode bending asymmetric ($E'$) quantum number. \\
$l_4$: Normal mode bending angular momentum quantum number.\\
$\Gamma_{\rm vib}$:   Vibrational symmetry index in \Dh{3}(M) \\
$K$:       Projection of $J$ on molecular symmetry axis.\\
$\Gamma_{\rm rot}$:   Rotational symmetry index in \Dh{3}(M). \\
$C_i$: Coefficient with the largest contribution to the $(J=0)$ contracted set; $C_i\equiv =1$ for $J=0$. \\
TROVE (local mode) quantum numbers: \\
$v_1$--$v_3$:  Stretching quantum numbers.\\
$v_4$, $v_5$:   Asymmetric bending quantum numbers.\\
$v_6$:  Inversion quantum number.\\
}
\end{table}

\begin{table}
\caption{Extract from the transitions file for the eXeL line list.  }
{\tt
\begin{tabular}{rrr}
\hline\hline
$f$ & $i$ & $A_{fi}$\\
\hline
        9135   &     4964  &  9.4529E-06  \\
        3483   &     2058  &  1.9377E-04  \\
        2590   &     4967  &  3.1507E-05  \\
        9141   &     4967  &  1.1550E-04  \\
        9142   &     1033  &  1.4600E-02  \\
        4975   &     9135  &  2.1565E-04  \\
        3484   &     7754  &  4.0709E-02  \\
        9142   &     4968  &  2.3899E-02  \\
        4979   &     2589  &  2.7283E-04  \\
        9147   &     4969  &  3.8512E-04  \\
\hline\hline
\end{tabular}
}
\label{t:trans}
\mbox{}\\
{$f$}: Upper state counting number.  \\
{$i$}: Lower state counting number. \\
$A_{fi}$: Einstein-A coefficient in s$^{-1}$.\\
\end{table}

The 5th columns of the \HTO\ States file contains the uncertainty \citep{jt804} of the corresponding term value (\cm),  estimated using the following conservative criterion:
$$
\sigma = 0.2\, n_1 + 0.2\, n_2 + 0.2\, n_3 + 0.2\, n_4 + 0.005 J (J+1).
$$
Only for the states replaced by empirical values the empirical uncertainties by \cite{09YuDrPe.H3O+} were used;
these uncertainties are all significantly smaller than estimated uncertainties of our computed levels.

The rotation-vibrational ground state $J=0$, $(n_1=0,n_2=0,n_3=0,n_4=0)$ has the symmetry $A'_1$ and therefore does not exist. The lowest existing ro-vibrational state is $J=1, K=1$ ($E''$) of $(0,0,0,0)$ with the difference of $hc\cdot$17.3803~\cm\ above the ground state. Following the same convention used for the BYTe line list for Ammonia \citep{jt500}, here we chose the zero-point-energy (ZPE) of the state  $(0,0,0,0)$, $J=0$, see  Table~\ref{t:obs-cacl}, which we estimated as $hc\cdot$7436.6~\cm\ relative to the minimum of the refined PES of \HTO. Therefore the line intensities as well as partition functions were computed using this convention.

\begin{figure}
\centering
\includegraphics[width=0.9\columnwidth]{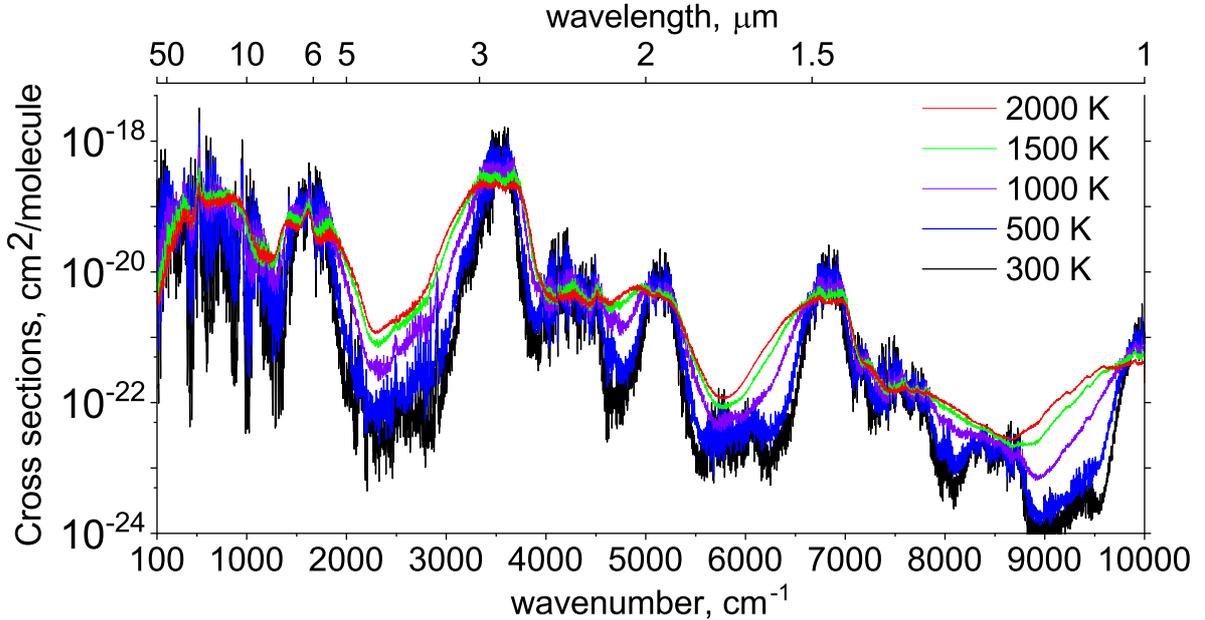}
\caption{Temperature dependence of the H$_3$O$^+$ absorption spectrum: the spectrum becomes flatter with increasing temperature. The spectrum was computed using the Gaussian line profile with HWHM of 1~\cm.}
\label{fig:Temp}
\end{figure}

\begin{figure}
\centering
\includegraphics[width=0.7\columnwidth]{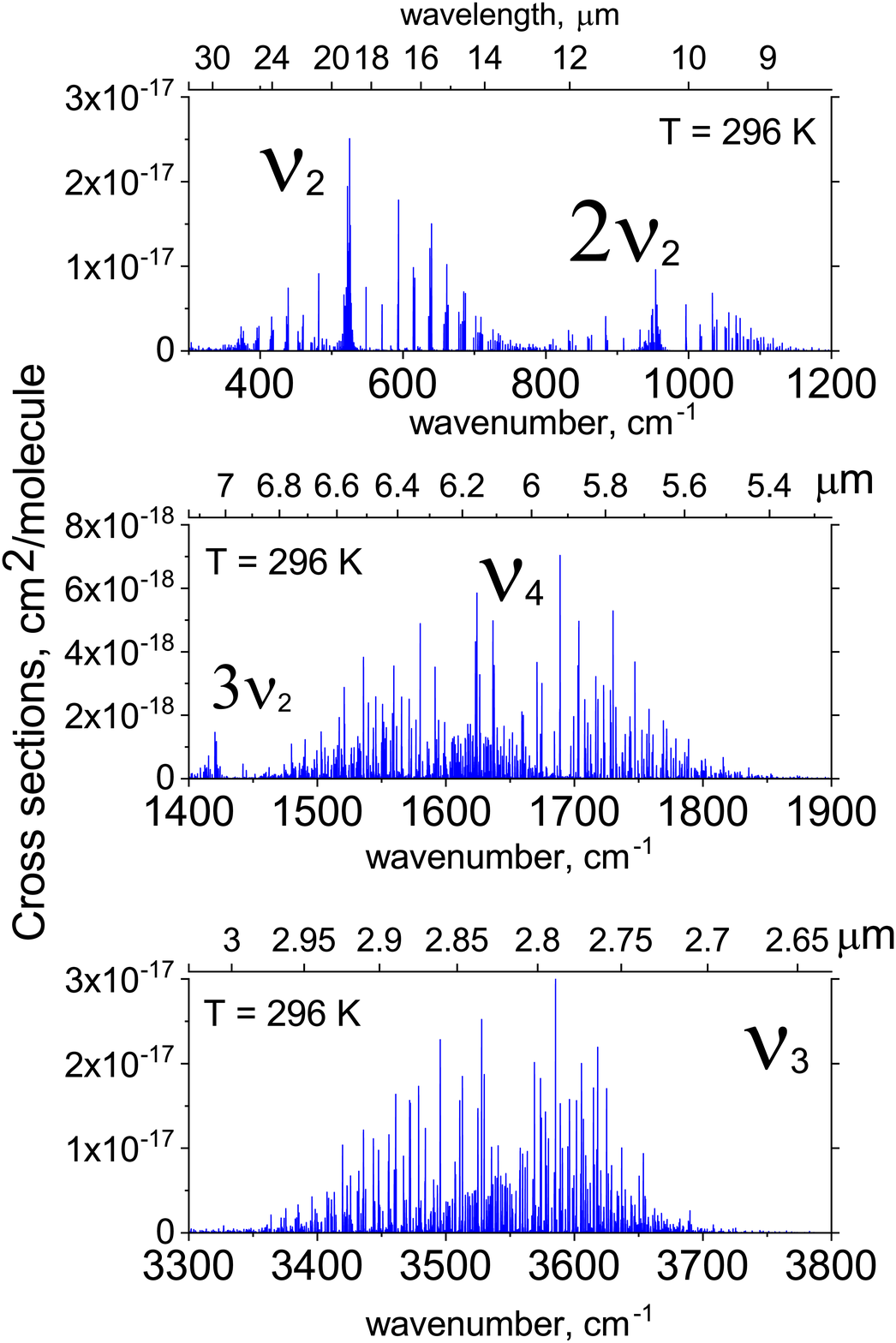}
\caption{Fundamental and overtone bands of H$_3$O$^+$ in absorption at $T=$ 296~K computed using the eXeL line list and the Doppler line profile.}
\label{fig:bands}
\end{figure}

An overview of the absorption spectra at a range of temperatures is shown in Fig.~\ref{fig:Temp}. The spectra were computed using the eXeL line list on a grid of 1~\cm\ assuming a Gaussian line profile of half-width-of-half-maximum (HWHM) of 1 \cm.
The strongest band is $\nu_3$ at 2.9~\um. Table~\ref{t:tm} lists vibrational transition moments for several strongest bands of \HTO\ computed using the eXeL line list and Fig.~\ref{fig:bands} illustrates five main fundamental and overtone bands at $T=$~296~K in absorption.

\begin{table}
    \centering
    \caption{Vibrational transition moments (Debye) $\bar\mu$ and band centers $\tilde\nu$ for selected bands of \HTO\ originated from the two components of the ground state, $0^+$ and $0^-$ and computed using eXeL.}
    \label{t:tm}
    \begin{tabular}{lrr}
 \hline\hline
 Band & \multicolumn{1}{c}{$\tilde\nu$ (\cm)} & \multicolumn{1}{c}{$\bar\mu$ (Debye)} \\
 \hline
 $  0^-            $&     55.403      &    1.4375   \\
 $  \nu_2^+-0^-    $&     525.717     &    0.7337   \\
 $  2\nu_2^-       $&     954.395     &    0.2888   \\
 $  3\nu_2^+ - 0^- $&    1421.231     &    0.1038   \\
 $  \nu_4^+        $&    1625.971     &    0.2307   \\
 $  \nu_4^- - 0^-  $&    1638.638     &    0.2241   \\
 $  2\nu_4^{0+}    $&    3240.946     &    0.0431   \\
 $  2\nu_4^{0-} -  $&    3267.694     &    0.0482   \\
 $  \nu_1^+ - 0^-  $&    3389.722     &    0.0505   \\
 $  \nu_1^-        $&    3491.340     &    0.0460   \\
 $  \nu_3^+        $&    3536.002     &    0.3326   \\
 $  \nu_3^- - 0^-  $&    3519.339     &    0.3274   \\
 \hline\hline
    \end{tabular}
\end{table}

\begin{figure}
\centering
\includegraphics[width=0.9\columnwidth]{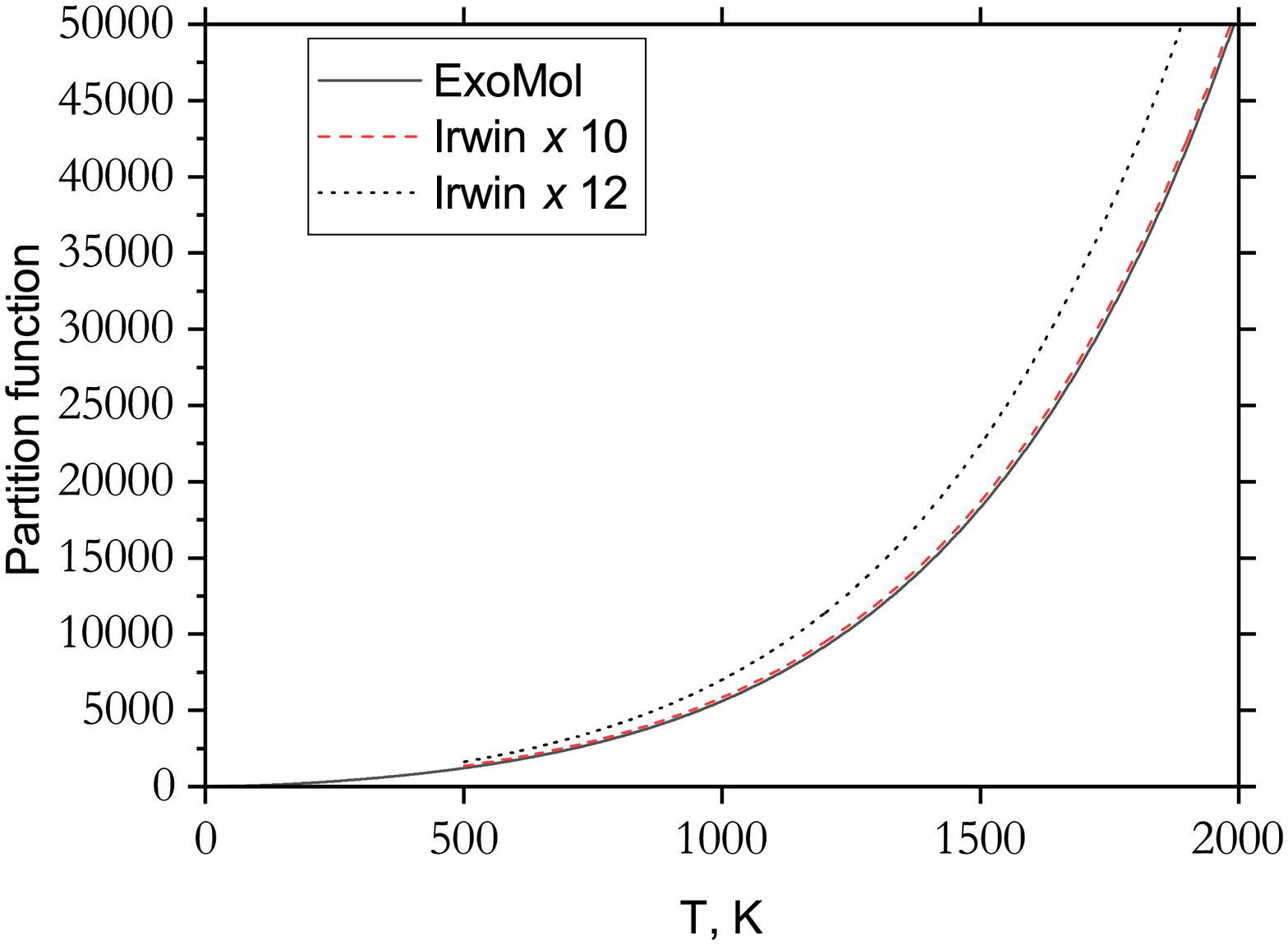}
\caption{Partition functions of \HTO\ computed using eXeL energies and constants provided by \citet{88Irwin}. }
\label{fig:pf}
\end{figure}

An eXeL partition function was computed on a 1~K grid of temperatures up to $T=1500$~K. Fig.~\ref{fig:pf} compares this partition function with that by \citet{88Irwin} produced for JANAF polyatomic molecules. The latter had to be multiplied by 10 in order to get best agreement with ExoMol which follows HITRAN's convention \citep{jt692} of using the full nuclear spin multiplicites.

\begin{figure}
\centering
\includegraphics[width=0.9\columnwidth]{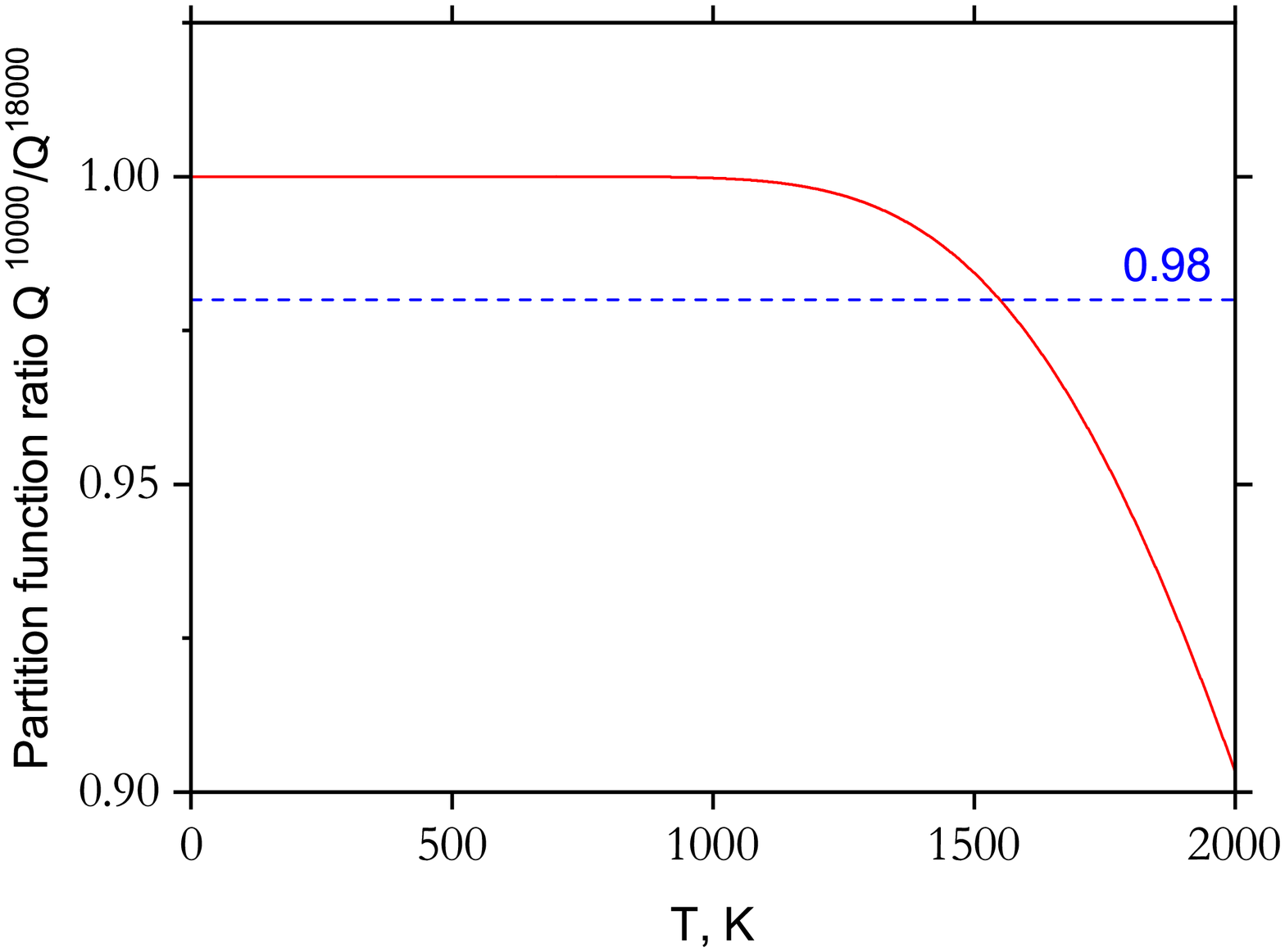}
\caption{Ratio of two partition functions of \HTO, $Q^{10000}$ (computed using all energies below $hc\cdot$10~000~\cm) and $Q^{18000}$ (computed  using all energies below $hc\cdot$18~000~\cm).}
\label{f:pf:ratio}
\end{figure}

In order to estimate the effect of the energy threshold of 10~000~\cm\ in the completeness of the line list for different ratio we have computed a partition function of \HTO\ using energies below 10~000~\cm, $Q^{10000}(T)$ and compared to that of the complete partition function $Q^{18000}(T)$ (here approximated by the energies below 18~000~\cm).  Figure~\ref{f:pf:ratio} shows a ratio $Q^{10000}(T)/Q^{18000}(T)$ of the partition functions. At $T=1550$~K the partition function of \HTO\ should be 98~\% complete.

\section{\HTO\ in planetary atmospheres and cool stars}

For many years \HTO\ has been considered as an important ion in the atmospheres of giant planets where O-rich materials are being deposited (see \citet{00MoBaxx}). In 2011, \citet{13ODStMe} detected a series of ``peaks and troughs'' in the pole-to-pole H$_3^+$ emission spectrum of Saturn, re-detecting these features in 2013 \citep{17ODMoCo}. These features corresponded to locations where magnetic field lines passing through the planet's rings connected to the upper atmosphere. \citet{17ODMoCo} explained the peaks as being formed because water-derived ions from the rings were soaking up electrons in the northern mid-latitude ionosphere, reducing the rate of H$_3^+$ dissociative recombination and consequently producing a relative local increase in this ion's density. At southern mid-latitudes, however, the influx of water was sufficiently large to overwhelm this effect and produce a local minimum of H$_3^+$ as a result of proton-hopping:
\begin{equation}
\rm{H}_3^+ + \rm{H}_2\rm{O} \rightarrow \rm{H}_2 + \rm{H}_3\rm{O}^+.
\end{equation}
Making use of modelling by \citet{15MoODMu}, \citet{19ODMoCo} deduced that the rings of Saturn would be fully eroded in between 168 and 1110 million years time at the current rate of water deposition.

A much larger equatorial mass influx from Saturn's rings, primarily composed of neutral nanograins, was discovered by the Cassini spacecraft during its end-of-mission proximal orbits \citep{18MiPeHa,18HsScKe,18WaPePe}.  Such a large mass loss from the rings could imply an even shorter lifetime, or perhaps a highly temporally variable process \citep{18PeWaMi}, though it is clear that deducing the lifetime of Saturn's rings from such limited measurements warrants caution (e.g., \citet{19CrChHs}).

\citet{18MoCrMu} have analysed data from the final orbits of the Cassini spacecraft and deduced that molecular ions with a mean mass of 11 Daltons dominate Saturn's lower ionosphere in the planet's equatorial regions, within the range derived from observations by the Cassini spacecraft \citep{18MoWaAn,18WaMoHa}.
The model of \citet{18MoCrMu} produces an \HTO\ density of 10$^9$ m$^{-3}$ at an altitude around 1500 km.

\begin{figure}
\centering
\includegraphics[width=0.7\columnwidth]{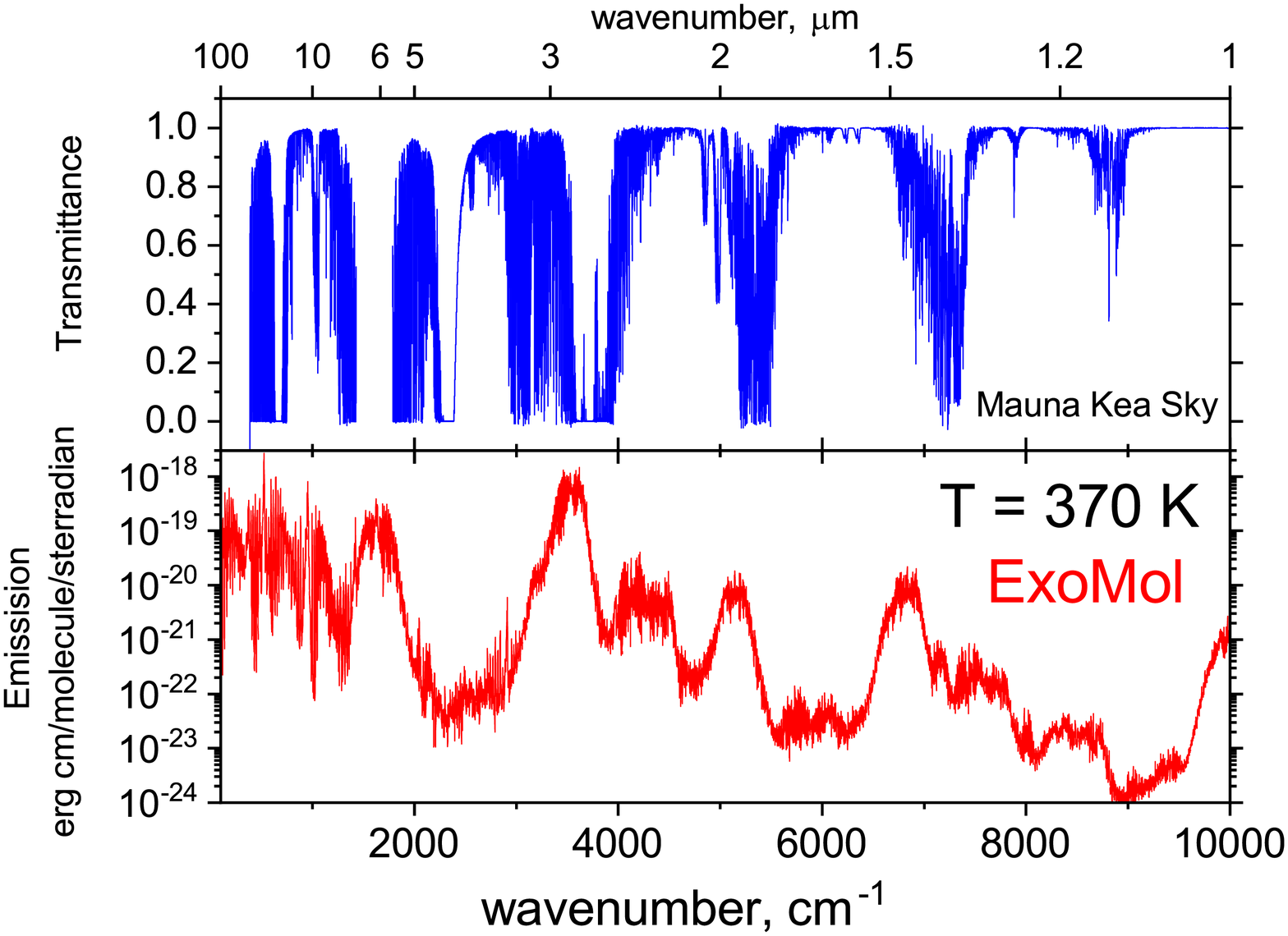}
\includegraphics[width=0.7\columnwidth]{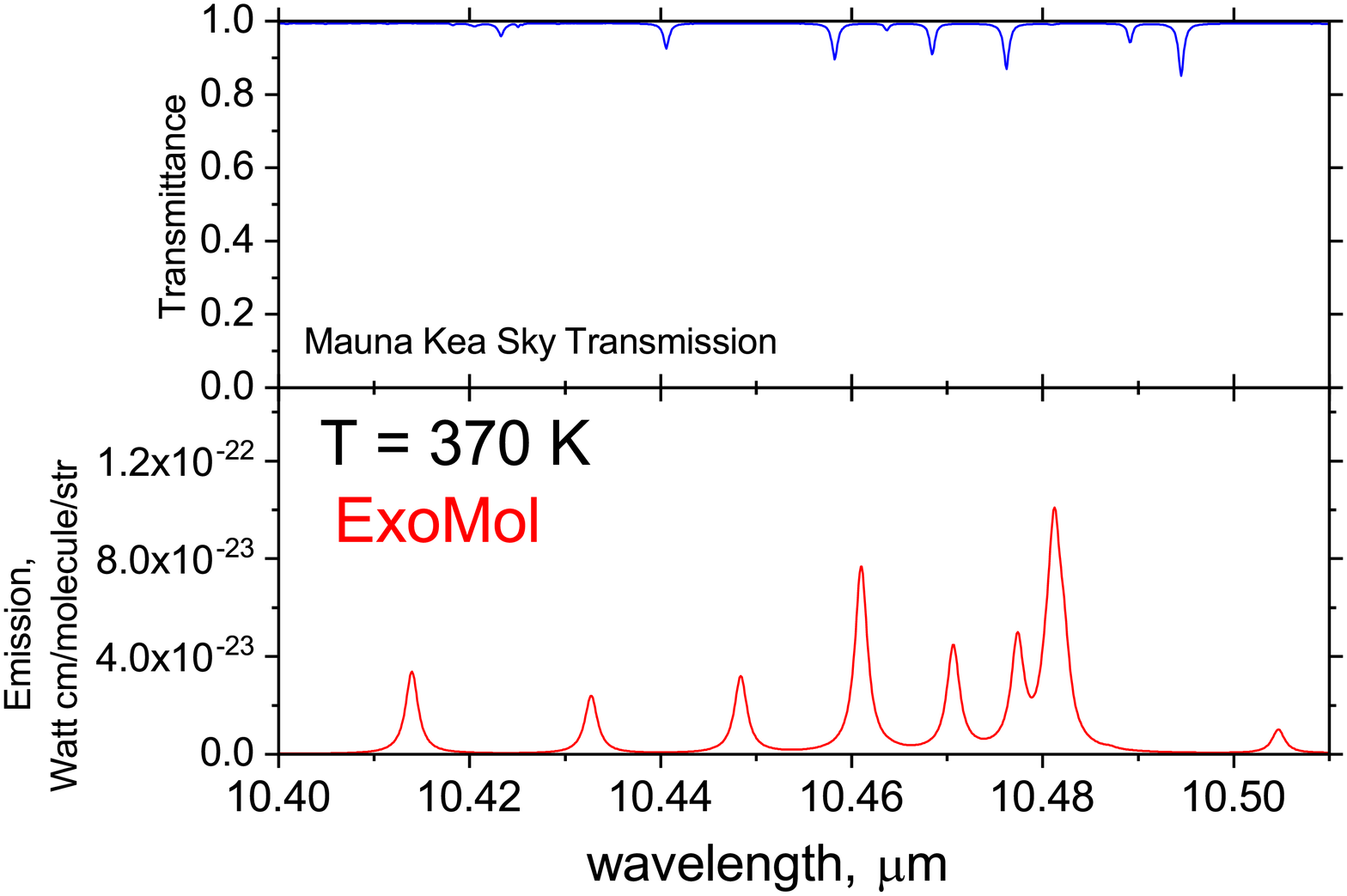}
\caption{Spectrum of \HTO\ at $T=370$~K (top: emission cross sections; bottom: emission line intensities) together with a spectrum of the Earth atmosphere at Mauna Kea (water vapor column 1.0~mm and air mass=1) \url{www.gemini.edu}. The \HTO\ spectrum was simulated assuming a Doppler line profile in air. The \HTO\ line positions were red-shifted by an equivalent of  20 km/s.}
\label{fig:Saturn}
\end{figure}

Figure~\ref{fig:Saturn} shows an \HTO\ emission spectrum of Saturn's equator, where \HTO\ is expected to be as large as $N$ $\sim$ 1.2$\times 10^{15}$ m$^{-2}$ \citep{18MoCrMu}, modelled using this line list. The temperature is assumed to be $T = $370 K, reasonable for Saturn's equatorial ionosphere \citep{18YeSeKo,20BrKoMu}. The possibility of making a detection from a ground-based infrared observatory is demonstrated by comparing this figure with a transmission spectrum of the terrestrial atmosphere at the summit of Maunakea, Hawai'i, using the data provided by the Gemini Observatory\footnote{\href{https://www.gemini.edu/sciops/telescopes-and-sites/observing-condition-constraints/ir-transmission-spectra}{{https://www.gemini.edu/sciops/telescopes-and-sites/observing-condition-constraints/ir-transmission-spectra}}}.

In particular, in Fig.~\ref{fig:Saturn}, we identify a spectral ``window'' around 10.40 to 10.50 $\mu$m where a blend of \HTO\ lines are clear of atmospheric absorption, even more so because we have included a red shift of Saturn's spectrum equivalent to 20 km/s, such as would be the typical case a few months past opposition as the planet recedes. This spectrum has been generated using a spectral resolving power $\lambda / \Delta \lambda = 85 000$, the resolving power of the TEXES mid-infrared spectrometer \citep{00LaRiGr} which is often used by telescopes belonging to the Mauna Kea Observatory group. Based on Figure \ref{fig:Saturn}, this spectral region is extremely promising
for the first detection of \HTO\ in a planetary atmosphere.

As well as the Solar System's giant planets, there is now considerable interest in determining the composition of giant exoplanet atmospheres and those of cool stars. Recently \citet{19HeRixx} have discussed the possibility of detecting \HTO\ in exoplanets and brown dwarf stars. They modelled the atmosphere of an M8.5 dwarf with an effective temperature of 2600~K. Their model indicated that the \HTO\ density is likely to be $\ge 10^{11}$ m$^{-3}$ throughout the pressure range from 1 bar to 1 $\mu$bar, and considerable proportion of the star's atmosphere. They concluded that this class of star could be a target for high-temperature \HTO\ emission in future studies: this could particularly be the case with the launch of the James Webb Space Telescope and its MIRI instrument (see, for example, \citet{20MaDeDi}). \citet{jt793} have also suggested that H$_3$O$^+$ could be detectable in the observational spectra of sub-Neptunes and proposed  H$_3$O$^+$ ions as potential biomarkers for Earth-like planets.

\section{Conclusion}

A new hot line list eXeL for \HTO\ is presented. The line list covers the wavenumber rage up to 10~000~\cm\ (wavelengths $>1$~\um) with the rotational excitation of $J=0-40$. The eXeL line list should be applicable for the temperatures up to 1500~K. There is evidence that this ion should be detectable in solar system gas
giants, exoplanets and brown dwarfs. The eXeL provides the spectroscopic data necessary for such
detections to be attempted.

Our line list for \HTO\ is aimed to help realistic simulations of absorption and emission properties of atmospheres of (exo-)planets and brown dwarfs as well as of cometary comae and interstellar clouds, their retrials and detections of \HTO.

ExoMol project originally concentrated on providing line lists for neutral molecules. At present
the database contains line lists for a number of ions of (possible) importance for studies of the early
Universe, namely HD$^+$ \citep{19AmDiJo}, HeH$^+$ \citep{jt347,19AmDiJo}, LiH$^+$ \citep{jt506}, and H$_3^+$ \cite{jt666}. For ions important in (exo-)planetary atmosphere the database so far only contains line lists
for H$_3^+$ and OH$^+$ \citep{MOLLIST,jt790}. The current \HTO\ line list represents an important
addition to this and we are in the process of adding other ions, starting with HCO$^+$.

The line lists can be downloaded from the CDS (\url{http://cdsweb.u-strasbg.fr/}) or from ExoMol (\url{www.exomol.com}) databases.

\section*{Acknowledgments}

This work was supported by the STFC Projects No. ST/M001334/1 and ST/R000476/1. The authors acknowledge the use of the UCL Legion High Performance Computing Facility (Legion@UCL) and associated support services in the completion of this work, along with the Cambridge Service for Data Driven Discovery (CSD3), part of which is operated by the University of Cambridge Research Computing on behalf of the STFC DiRAC HPC Facility (www.dirac.ac.uk). The DiRAC component of CSD3 was funded by BEIS capital funding via STFC capital grants ST/P002307/1 and ST/R002452/1 and STFC operations grant ST/R00689X/1. DiRAC is part of the National e-Infrastructure.

\section*{Data availability statement}

Full data is made available. The line lists can be downloaded from the CDS (\url{http://cdsweb.u-strasbg.fr/}) or from ExoMol (\url{www.exomol.com}) databases. The following files are available as supplementary information:
\begin{tabbing}
H3Op\_PES\_refined.inp \= asdaddaaa \= \kill
H3Op\_PES\_refined.inp  \>  Input file for HO3p\_PES.f90 containing the potential parameters defining refined PES of H$_3$O$^+$\\
HO3p\_PES.f90   \>          Fortran 90 routine for calculating potential energy values in combination with \\
\> the input file H3Op\_PES\_refined.inp\\
H3Op\_DMS.inp    \>         Input file for H3Op\_DMS.f90 containing dipole moment parameters defining ab initio DMS of H$_3$O$^+$\\                      
H3Op\_DMS.f90     \>        Fortran 90 routine for calculating dipole momment values in combination with the input file H3Op\_DMS.inp\\
\end{tabbing}

\label{lastpage}

\bibliographystyle{mn2e}

\end{document}